\documentclass[amsmath,amssymb,aps,prl,twocolumn,showpacs,superscriptaddress]{revtex4-1}

\usepackage{graphicx}
\usepackage[colorlinks=true,linkcolor=blue,citecolor=blue,urlcolor=blue]{hyperref}
\usepackage{color}

\newcommand{\om}{\omega}

\newcommand{\ii}{i}
\newcommand{\ee}{e}
\newcommand{\dd}{d}

\newcommand{\ev}[1]{\langle #1 \rangle}
\newcommand{\evs}[2]{\langle #1 \rangle_{#2}}
\newcommand{\tr}{\mbox{tr}}
\newcommand{\comm}[2]{[ #1 , #2 ]}
\newcommand{\eq}[1]{(\ref{eq:#1})}

\begin{document}

\title{Entangled phonons in atomic Bose--Einstein condensates}

\author{Stefano Finazzi}
\email{stefano.finazzi@univ-paris-diderot.fr}
\affiliation{Laboratoire Mat\'eriaux et Ph\'enom\`enes Quantiques, Universit\'e Paris Diderot-Paris 7 and CNRS, B\^atiment Condorcet, 10 rue Alice Domon et L\'eonie Duquet, 75205 Paris Cedex 13, France.}
\affiliation{INO-CNR BEC Center and Dipartimento di Fisica, Universit\`a di Trento, via Sommarive 14, 38123 Povo--Trento, Italy}
\author{Iacopo Carusotto}
\email{carusott@science.unitn.it}
\affiliation{INO-CNR BEC Center and Dipartimento di Fisica, Universit\`a di Trento, via Sommarive 14, 38123 Povo--Trento, Italy}

\begin{abstract}
We theoretically study the entanglement between phonons spontaneously generated in atomic Bose-Einstein condensates by analog Hawking and dynamical Casimir processes. The quantum evolution of the system is numerically modeled by a truncated Wigner method based on a full microscopic description of the condensate and state non-separability is assessed by applying a generalized Peres-Horodecki criterion. The peculiar distribution of entanglement is described in both real and momentum spaces and its robustness against increasing initial temperature is investigated. Viable strategies to experimentally detect the predicted phonon entanglement are briefly discussed.
\end{abstract}

\date{\today}
\maketitle

Pairs of correlated photons are a crucial element of many quantum optical experiments, from Hong-Ou-Mandel two-photon interference~\cite{HOM}, to fundamental tests of Bell inequalities~\cite{Aspect}, to linear optical schemes for quantum information processing~\cite{raussendorf2001}. Among the most common strategies to create such pairs there are atomic two-photon decays and spontaneous parametric down-conversion in coherently pumped nonlinear crystals.

Following the dramatic advances in cooling and manipulating atomic samples, many groups have recently demonstrated entanglement effects using atomic matter waves: Among the most striking results, we can mention non-classical violations of the Cauchy-Schwartz inequalities in the product of atomic collisions~\cite{schmiedmayer,chris} and noise reduction in atomic interferometry experiments using squeezed matter fields~\cite{yun,homodyne}. 

A most exciting challenge is now to extend quantum optical concepts to the {\em quantum hydrodynamic} context, so to generate and manipulate non-classical states of the hydrodynamical degrees of freedom of a macroscopic fluid~\footnote{Note that the term {\em quantum hydrodynamics} is used here in a somehow stricter sense than in most many-body literature where it broadly refers to generic hydrodynamic effects in a quantum fluid.}.
Among the first proposals in this direction, the pioneering work~\cite{unruh81} anticipated that quantum hydrodynamical fluctuations in a moving fluid are converted at a black hole sonic horizon into pairs of propagating phonons via a mechanism analogous to the Hawking effect of gravitational physics~\cite{hawkingnat}.

In the following years, this analogy between phonons propagating on top of a fluid and quantum fields living on a curved space-time has experienced impressive developments~\cite{lr}. In particular, atomic Bose-Einstein condensates (BEC) have turned out to be a most promising platform to observe the analog Hawking radiation (HR)~\cite{garay,garaypra} as well as the quantum emission of opposite momentum phonon pairs when the speed of sound in a homogeneous fluid is made to rapidly vary in time~\cite{barcelo,fedichev2003,kramer2005,jain2007,iacopoEPJD}. This latter effect can be considered as a condensed-matter analog of the dynamical Casimir effect (DCE)~\cite{casimir,lambrecht2005} or of a quickly expanding universe~\cite{parker,inflation}.
Experimental  investigations along these lines are presently very active: a sonic horizon in a flowing atomic BEC was realized~\cite{steinhauer} and the emission of classically correlated pairs in a temporally modulated BEC was observed~\cite{chrisdyn}.

In spite of these on-going experimental activities, only a limited attention~\cite{reznik,zapata,nott} has been paid so far to the quantum entanglement aspects of the analog DCE and HR phonon emission processes. In the present Letter we theoretically investigate the spatial and spectral entanglement features displayed by the emitted phonons and we discuss strategies to detect this entanglement in realistic experiments using atomic BECs. On one hand, detection of such entanglement would provide an ultimate proof of the quantum nature of the analog DCE and HR emissions.
On the other hand, these same processes will be of great utility as flexible sources of entangled phonons to be used in quantum hydrodynamical experiments.

\paragraph{The generalized Peres--Horodecki (gPH) criterion.---}%
Among the many available criteria to assess whether a given state of a bipartite quantum system is entangled or not~\cite{horodeckirev}, in this Letter we shall adopt the one originally proposed by Peres and Horodecki in~\cite{peres,horodecki} and later extended~\cite{duanetal,simon} to the continuous variable case of present interest. Consider two generic degrees of freedom described by the operators $\hat q_j$ and $\hat p_j$ ($j=1,2$), satisfying the usual canonical commutation relations $\comm{\hat q_j}{\hat q_k}=\comm{\hat p_j}{\hat p_k}=0$, $\comm{\hat q_j}{\hat p_k}=\ii\,\delta_{jk}$.  
These operators can be grouped into a single 4-component vector $\hat X\equiv(\hat q_1,\hat p_1,\hat q_2,\hat p_2)$ of expectation value $\bar X=\evs{\hat X}{\hat\rho}=\tr(\hat X\hat\rho)$ and covariance matrix $V_{\alpha\beta}=\frac{1}{2}\evs{\Delta\hat X_\alpha\Delta\hat X_\beta+\Delta\hat X_\beta\Delta\hat X_\alpha}{\hat\rho}$, where $\Delta\hat X=\hat X-\ev{\hat X}$.

Following Simon~\cite{simon}, the gPH criterion states that if a state is separable then the following quantity
\begin{multline}\label{eq:P}
 \mathcal{P}\equiv\det A\det B+\left(\frac{1}{4}-|\det C|\right)^2\\-\tr[AJCJBJC^TJ]-\frac{1}{4}(\det A+\det B)
\end{multline}
is positive, where $A, B, C$ are $2\times2$ submatrices of the $4\times 4$ covariance matrix $V$ and $J$ is the $2\times2$ simplectic matrix
\begin{equation}
 V=
 \begin{pmatrix}
  A & C\\ C^T & B
 \end{pmatrix},
 \quad
 J=
 \begin{pmatrix}
  0 & 1\\-1 & 0
 \end{pmatrix}.
\end{equation}
%

\paragraph{Construction of phonon operators.---}%
The first step to assess entanglement between phonons propagating on top of an atomic BEC consists of identifying the relevant degrees of freedom and constructing the corresponding operators $\hat q_j$ and $\hat p_j$ or, equivalently, the associated bosonic annihilation operators $\hat a_{j}=(\hat q_j+\ii\, \hat p_j)/\sqrt{2}$, satisfying $\comm{\hat a_j}{\hat a_k}=0$ and $\comm{\hat a_j}{\hat a_k^\dagger}=\delta_{jk}$.
In order to capture entanglement features in both real and momentum spaces, it is convenient to define phonon wavepacket operators $ \hat a_{k,x_0}$ by projecting the full atomic field $\hat\Psi(x)$ onto a localized phonon wavepacket
\begin{equation}\label{eq:ap}
 \hat a_{k,x_0}=\int\dd x\, f_{k,x_0}^*(x)\left[u_{k}\,\hat{{\Phi}}(x)-v_{k} \hat{{\Phi}}(x)^\dagger\right].
\end{equation}
with a Gaussian envelope
\begin{equation}
 f_{k,x_0}(x)=\left(\pi\sigma^2\right)^{-1/4}\ee^{-(x-x_0)^2/2\sigma^2}\ee^{\ii k (x-x_0)}
 \label{eq:gauss}
\end{equation}
centered at $x_0$, of width $\sigma$ and carrier wavevector $k$.
The exponential factor in the definition $\hat{{\Phi}}(x)\equiv e^{-i(K x- \Omega t) }\,\hat{{\Psi}}(x)$ of the normalized atomic field takes into account the space- and time-dependence of the condensate phase with wavevector $K$ and frequency $\Omega=\hbar K^2/2m + gn$. The Bogoliubov coefficients $u_k$ and $v_k$ in \eqref{eq:ap} serve to extract the phonon amplitude from the full atomic field. As usual~\cite{bectextbook}, they are defined by $u_k\pm v_k=[{\varepsilon_k}/({\varepsilon_k+2\mu})]^{\pm 1/4}$
in terms of the kinetic and interaction energies $\varepsilon_k={\hbar^2k^2}/{2m}$ and $\mu=gn$, where $m$ is the atomic mass, $n$ the one-dimensional density, and $g$ the effective one-dimensional interaction constant~\cite{iacopo}.

In the following, we will consider measurements performed with a pair of wavepacket operators $\hat a_1=\hat a_{k_1,x_1}$ and $\hat a_2=\hat a_{k_2,x_2}$, with wavevectors $k_{1,2}$ and centered at $x_{1,2}$. As a first step, one must ensure that $\hat a_{1,2}$ satisfy with good accuracy the proper bosonic commutation rules. 
Using the canonical commutation relation for $\hat\Psi$ and assuming $\sigma |k_{1,2}|\gg 1$, so to suppress overlap with the condensate mode~\cite{bectextbook}, one finds
\begin{align}
 \comm{\hat a_{k_1,x_1}}{\hat a_{k_2,x_2}}&=(v_{k_1} u_{k_2}-u_{k_1} v_{k_2})\ee^{\ii(x_1-x_2)(k_1-k_2)/2}\nonumber\\
 &\quad\times\ee^{-(x_1-x_2)^2/4\sigma^2}\ee^{-\sigma^2(k_1+k_2)^2/4},\\
 \comm{\hat a_{k_1,x_1}}{\hat a_{k_2,x_2}^\dagger}&=(u_{k_1} u_{k_2}-v_{k_1} v_{k_2})\ee^{\ii(x_1-x_2)(k_1+k_2)/2}\nonumber\\
 &\quad\times\ee^{-(x_1-x_2)^2/4\sigma^2}\ee^{-\sigma^2(k_1-k_2)^2/4}.
\end{align}
As expected on physical grounds, all commutators
vanish as soon as the two points $x_{1,2}$ are separated by a distance $|x_1-x_2|\gg \sigma$ and $\sigma |k_{1,2}|\gg 1$. Under this latter condition, $\comm{\hat a_{k_1,x_1}}{\hat a_{k_2,x_2}}\simeq 0$ even when the two points are close ($|x_1-x_2|\lesssim \sigma$). On the other hand, the bosonic nature of each operator $\hat a_{k,x_0}$ is validated by the fact that $\comm{\hat a_{k,x_0}}{\hat a_{k',x_0'}^\dagger}$ accurately recovers the desired value $1$ for wavepackets overlapping in both real ($|x_0-x_0'|\ll \sigma$) and momentum ($\sigma |k-k'| \ll 1$) spaces.

\paragraph{Numerical calculation of the gPH function.---}%

To calculate the covariance matrix $V$ and then the gPH function $\mathcal{P}$ we use the truncated Wigner method~\cite{steel,sinatra} in which the expectation value of any symmetrized operator is obtained as the statistical average of the corresponding classical quantity over the Wigner distribution. In our calculations this distribution is numerically sampled along the lines of~\cite{iacopo} by following in time the Gross--Pitaevskii evolution of $N_r$ classical wavefunctions $\Psi^{j=1,\ldots ,N_r}(x,t)$ starting from suitably chosen initial conditions $\Psi^j(x,0)$ that reproduce either the the ground state or a thermal distribution. For instance, the expectation value of the operator $\hat a_{k,x_0}$ is obtained as the statistical average 
\begin{equation}
 \ev{\hat a_{k,x_0}}=\frac{1}{N_r}\sum_{j=1}^{N_r} \alpha^j_{k,x_0}
\end{equation}
of the corresponding classical quantity as extracted from each realization $\Psi^{j}(x)$ via the classical counterpart of \eqref{eq:ap},
\begin{equation}\label{eq:alphap}
 \alpha^j_{k,x_0}=\int\dd x\, f_{k,x_0}^*(x)\left[u_{k}{\Phi}^j(x) -v_{k}{\Phi}^j(x)^*\right].
\end{equation}
Analogously, all the elements of the covariance matrix involved in the gPH function $\mathcal{P}$ via Eq.~\eqref{eq:P} are obtained as the statistical average of the products of the corresponding classical amplitudes \eq{alphap}.

\paragraph{Analog DCE in temporally modulated condensates.---}%

Inspired by the experiment~\cite{chrisdyn}, as a first application of the gPH criterion we study the entanglement between the phonons spontaneously created in a spatially homogeneous condensate by a sudden temporal variation of the speed of sound $c$ from $c_-$ at $t=0^-$ to $c_+$ at $t=0^+$, obtained through a sudden modulation of the atomic interaction energy. This process is a condensed-matter analog of the dynamical Casimir effect or of the spontaneous particle creation in a quickly expanding (or contracting) universe~\cite{barcelo,fedichev2003,jain2007,iacopoEPJD}. A first study of entanglement for this configuration has recently appeared in~\cite{nott} but is restricted to correlations in wavevector space. Here we extend the discussion by unveiling peculiar structures in the space-wavevector pattern of entanglement.

\begin{figure}
 \includegraphics{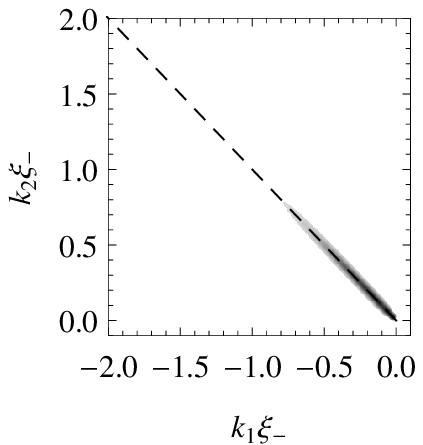}
 \includegraphics{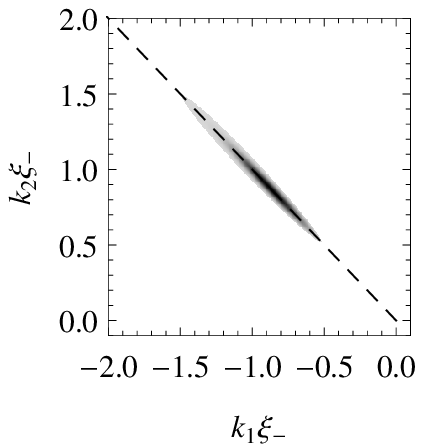}
 \includegraphics{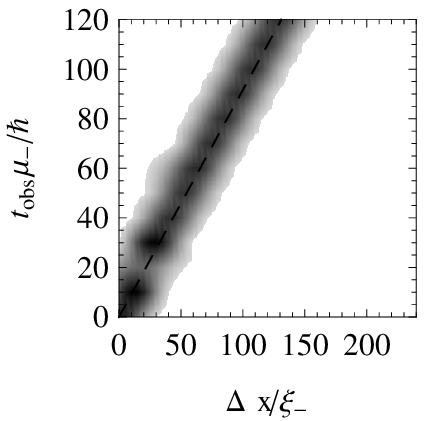}
 \includegraphics{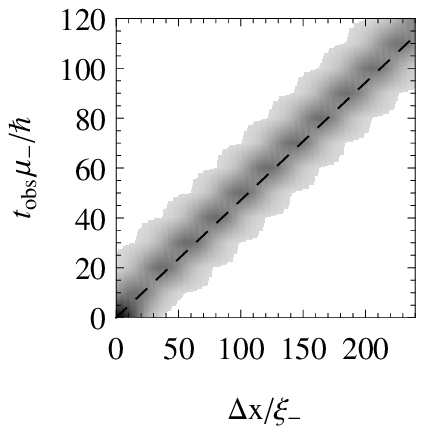}
 \caption{\label{fig:dce} Entanglement between dynamical Casimir phonons. Upper panels: Region of the $k_1$--$k_2$ momentum plane where $\mathcal{P}<0$ (darker gray, larger $|\mathcal{P}|$) for measurement wavepackets centered at distances  $\Delta x=|x_1-x_2|=120\xi_-$ (left) and $240\xi_-$ (right) with $\sigma=35\xi_-$. The dashed lines indicate the opposite momentum condition $k_2=-k_1$. Lower panels: Same plot in the $\Delta x$--$t_{\rm obs}$ plane for wavepackets centered at wavevectors $k_2=-k_1=0.25/\xi_-$ (left) and $1/\xi_-$ (right) with $\sigma=15\xi_-$. Dashed lines indicate the ballistic $t_{\rm obs}=\Delta x/2|v^{\rm gr}(k_{1,2})|$ condition.
All plots are for $t_{\rm obs}=120\hbar/\mu_-$ starting from zero initial temperature.
}
\end{figure}

In the upper panels of Fig.~\ref{fig:dce} we display the region in the $k_1$--$k_2$ momentum plane where the gPH function $\mathcal{P}$ is negative. In agreement with the translational invariance of the system, in both panels entanglement is visible only between phonons with opposite wavevectors $k_1=-k_2$ [up to an uncertainty $1/\sigma$ set by the wavepacket size $\sigma$]. In the two panels, the numerical observation is performed at the same time $t=t_{\rm obs}=120\hbar/\mu_-$ but for different values of the distance $\Delta x=x_2-x_1=120\xi_-$ (left) and $240\xi_-$ (right), $\xi_-=\hbar/mc_-$ being the initial healing length. In analogy with the density correlations discussed in~\cite{iacopoEPJD,angus}, entanglement along the $k_1=-k_2$ line is strongest at the momentum value $|k_{1,2}|=\bar{k}$ satisfying the ballistic condition $2 v^{\rm gr}(\bar{k}) t_{obs}=\Delta x$, where $v^{\rm gr}(k)=d\om/dk$ is the group velocity of phonons of wavevector $k$, as derived by the Bogoliubov dispersion $\om^2= c^2k^2+{\hbar^2k^4}/{4m^2}.$

This peculiar structure of the entanglement pattern is further illustrated in the lower panels of Fig.~\ref{fig:dce}, where entanglement is plotted in the $\Delta x$--$t_{\rm obs}$ plane for two given values of the wavevectors $|k_{1,2}|=\tilde{k}$. As expected, entanglement is concentrated within a distance $\sigma$ from the (dashed) straight line $t_{obs}=\Delta x/2|v^{\rm gr}(k)|$ whose slope is determined by the selected phonon wavevector $k=\tilde{k}$.

\paragraph{Analog HR from sonic horizons.---}%

\begin{figure}
 \includegraphics{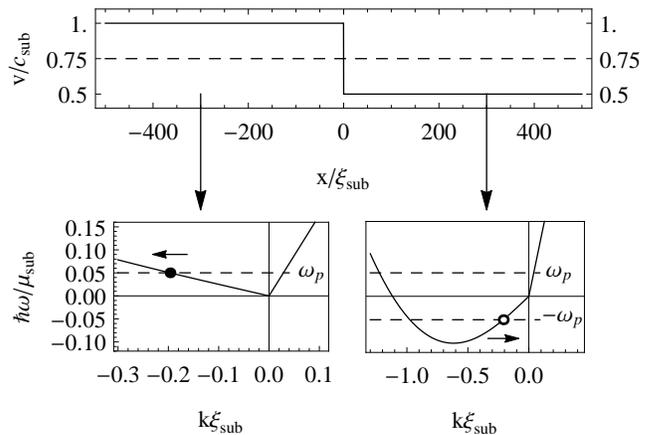}
 \caption{\label{fig:disp} Flowing condensate showing a sonic horizon at $x=0$.
Upper panel: flow velocity (dashed line) and speed of sound (solid line) profiles. Lower panels: Bogoliubov dispersion relation of phonons in the subsonic (left) and supersonic (right) regions. Open and closed dots indicate the modes into which the dominant Hawking emission occurs.
}
\end{figure}

As a second application, we study the case of a flowing condensate showing a sonic black hole (BH) horizon. An example of such a configuration is sketched in the upper panel of Fig.~\ref{fig:disp}. A condensate with a homogeneous density $n(x)=n_0$ flowing with a uniform velocity $v(x)=v_0>0$ directed in the rightward direction is considered. The sonic horizon is created by a suitable dependence of the interaction constant $g(x)$ and of the external potential $V(x)$~\cite{iacopo}. In the upstream (downstream) region, corresponding to the exterior (interior) of the BH, the flow is subsonic (supersonic) $c(x<0)=c_{\rm sub}>v_0$ [$c(x>0)=c_{\rm sup}<v_0$], while the point $x=0$, where $c(x)=v_0$, behaves as the sonic BH horizon.

Along the lines of~\cite{macherbec,pavloff}, the physical origin of the HR can be understood from the Bogoliubov dispersion relation $(\om-v_0 k)^2= c^2k^2+{\hbar^2k^4}/{4m^2}$ in the laboratory frame, which is shown in the lower panels of Fig.~\ref{fig:disp} for the subsonic (left) and supersonic (right) regions, respectively. The most salient feature is the presence in the super-sonic region (right panel) of modes with negative energy. As a result, pairs of phonons can be spontaneously created at no energy cost. The dominating process is the HR: The emission of a positive energy phonon of frequency $\omega=\omega_p$, propagating away from the BH (closed dot), is compensated by the simultaneous emission of a negative energy phonon of frequency $\omega=-\omega_p$, falling inside the BH (open dot). 
As a result, the energy conservation condition imposes a relation between the momenta $k_{1,2}$ of the two emitted phonons $\omega_{\rm sub}(k_1)=-\omega_{\rm sup}(k_2)$ in the sub- and super-sonic regions, respectively.

\begin{figure}
 \includegraphics{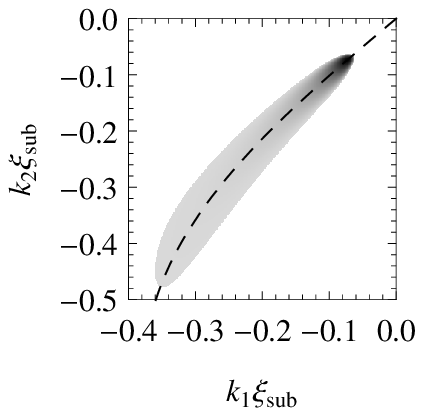}
 \includegraphics{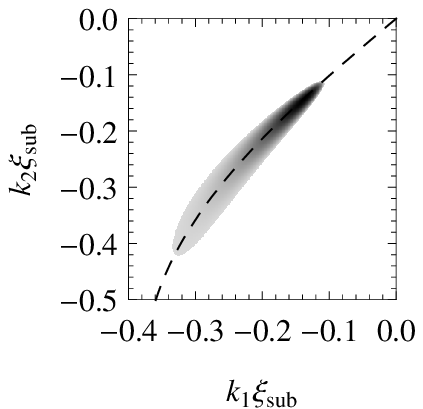}
\includegraphics{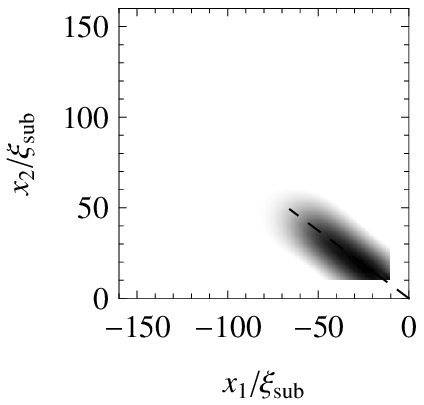}
\includegraphics{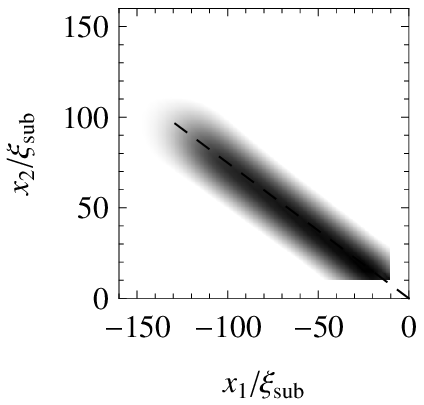}
 \caption{\label{fig:bhk} 
 Entanglement between Hawking phonons. Upper panels: region of the $k_1$--$k_2$ momentum plane where $\mathcal{P}<0$ (darker gray, larger $|\mathcal{P}|$) is negative for initial temperatures $k_B T/\mu_{\rm sub}=0$ (left) and $0.1$ (right) and observation time $t_{\rm obs}=480\hbar/\mu_{\rm sub}$. The measurement wavepackets are located at $x_1=-61\xi_{\rm sub}$ and $x_2=59\xi_{\rm sub}$ with $\sigma=35\xi_{\rm sub}$ so to optimize the entanglement signal. Dashed lines indicate the energy conservation condition (see text). Lower panels: Same plot in the $x_1$--$x_2$ plane for $t_{\rm obs}\mu_{\rm sub}/\hbar=240$ (left) and $480$ (right).  Wavepackets are centered at $k_2=k_1=-0.25/\xi_{\rm sub}$ with $\sigma=15\xi_{\rm sub}$. Initial temperature $T=0$. The dashed lines indicate the ballistic condition (see text). For all panels, the velocity profile is the one shown in Fig.~\ref{fig:disp}.}
\end{figure}

This feature is illustrated in the upper panels of Fig.~\ref{fig:bhk} where entanglement is plotted in the $k_1$--$k_2$ momentum plane for two different values of the initial temperature $k_B T/\mu_{\rm sub}=0$ (left) and $0.1$ (right); the spatial positions $x_{1,2}$ have been taken on the two sides of the horizon. As expected, entanglement is present for pairs of phonons of wavevectors $k_{1,2}$ within a distance $1/\sigma$ from the dashed line indicating the energy conservation condition. This result is the quantum entanglement counterpart of the momentum space correlations predicted in~\cite{Larre:PhD} and the sub-Poissonian features predicted in~\cite{zapata}.
In analogy with the density correlations~\cite{balbinot,iacopo}, entanglement is strongest along the energy conservation line when the ballistic condition $x_1/v_{\rm sub}^{\rm gr}(k_1)=x_2/v_{\rm sup}^{\rm gr}(k_2)$ is satisfied, meaning that the propagation time from the horizon is the same on the two sides.

As temperature increases, entanglement globally decreases at all momenta and starts disappearing from the low and high momentum regions. In the low momentum region, it is destroyed by the large initial thermal occupation of modes, in agreement with the analogous prediction for the DCE case~\cite{nott}. In the high momentum region, instead, the already very low Hawking emission rate gets easily blurred by the finite momentum resolution $1/\sigma$ of our measurement scheme~\footnote{We have validated this hypothesis by repeating the analysis with a smaller $\sigma$: As expected (not shown), the entanglement band in the $k_1$--$k_2$ plane becomes proportionally wider and the entanglement of high momentum modes disappears at lower temperatures.}. Eventually, entanglement is completely absent for $k_B T/\mu_{\rm sub}\gtrsim0.2$.

Finally, the above mentioned guess on the spatial structure of the entanglement pattern in the $x_1$--$x_2$ plane is confirmed in the lower panels of Fig.~\ref{fig:bhk}. Wavevectors $k_1=k_2=-0.25 /\xi_{\rm sub}$ approximately satisfying the energy conservation condition are chosen. As expected, entanglement is concentrated in the vicinity of the straight dashed lines indicating the ballistic condition.
As phonons start being steadily created since the formation of the horizon at $t=0$, entanglement is restricted to the accessible region $x_1>v^{\rm gr}_{\rm sub}(k_1)t_{\rm obs}$ and $x_2<v^{gr}_{\rm sup}(k_2)t_{\rm obs}$: this important feature is easily visible by comparing in the lower panels the length of the entanglement tongue at two different times $t_{\rm obs}\mu_{\rm sub}/\hbar=240$ (left) and $=480$ (right).

\paragraph{Experimental remarks ---}

An experimental measurement of the $\mathcal P$ function requires an efficient way of measuring the covariance matrix $V$ of the phonon quadratures. This route was recently followed to assess quantum correlations in the microwave DCE emission from modulated circuit-QED devices~\cite{delsing,paraoanu}. As a complete discussion of a specific experimental protocol goes beyond the scope of this Letter, we restrict ourselves to illustrate the feasibility of our proposal by sketching a couple of possible experimental strategies.

One possibility is to perform a preliminary phonon evaporation stage to map the phonon field onto a bare atomic field~\cite{tozzodalfovo,chrisdyn}, and then to perform a homodyne detection of the non-condensed atoms~\cite{homo1,homo2,review}. This technique requires a very precise control of the relative phase of the condensate and the matter wave local oscillators.

An alternative, possibly easier possibility consist of measuring the component of the atomic density fluctuation that corresponds to the phonon mode of interest. As it is detaled in the Supplemental Material, this may be done, e.g., by measuring the frequency shift of the optical modes of a cavity enclosing the condensate as in the ETH experiments of~\cite{esslingerscience,BECincavity,cavity_review}. The wavevectors of the phonon quadrature of interest are selected by the spatial structure of the cavity modes. Thanks to the non-destructive nature of such measurements, a proper series of measurements at different times provides all the elements of the covariance matrix $V$, from which the gPH function ${\mathcal P}$ can be directly estimated.

\paragraph{Conclusions.---}

In this Letter we have theoretically studied the quantum entanglement of phonons spontaneously generated in atomic Bose-Einstein condensates by processes analogous to the dynamical Casimir effect and the Hawking radiation. The experimental detection of such entanglement would be the smoking gun of the quantum origin of these phenomena in the amplification of zero-point fluctuations. In the future, we expect that these novel excitation scheme may become the phonon analog of parametric downconversion process in photonics, that is a flexible source of entangled phonons to be used in quantum hydrodynamical experiments. 

\acknowledgments{This work has been supported by ERC through the QGBE grant and by Provincia Autonoma di Trento. We are grateful to R. Balbinot, D. Gerace, and R. Parentani for continuous stimulating exchanges.}

\bibliography{correlations}

\newcommand{\Nq}[2]{{\hat N_{k_0}\left(#1,#2\right)}}
\newcommand{\pit}{\frac{\pi}{2}}
\newcommand{\pif}{\frac{\pi}{4}}
\newcommand{\Xq}{\hat q_{k_0}}
\newcommand{\Xmq}{\hat q_{-{k_0}}}
\newcommand{\Pq}{\hat p_{k_0}}
\newcommand{\Pmq}{\hat p_{-{k_0}}}

\clearpage

\section{SUPPLEMENTAL MATERIAL}

\section{Quantitative remarks on the quadratures and the gPH function $\mathcal{P}$ }

\begin{figure}[b]
\includegraphics{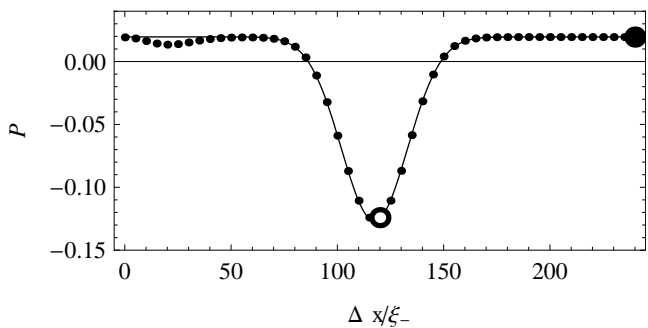}
\caption{\label{fig:cut}$\mathcal{P}$ as a function of the distance  $\Delta x$ between the centers of two wavepackets with opposite wavevectors $k_{2}=-k_{1}=0.25/\xi_{-}$ and spatial width $\sigma=15\xi_{-}$, at $t_{\rm obs}=110\,\hbar/\mu_{-}$ after the sudden temporal modulation originating the analog DCE (see left lower panel of Fig.~1 of the Main Text). Solid curve: best-fit of the gaussian function of Eq.~\eqref{eq:fit}; values of the parameters in Eq.~\eqref{eq:fitpar}.}
\end{figure}

In this section, we provide more quantitative insight on the behavior of the function $\mathcal{P}$. In Fig.~\ref{fig:cut}, we plot the numerical values of $\mathcal{P}$ (dots) as a function of $\Delta x$ for a cut at constant time $t_{\rm obs}$ of the left lower panel of Fig.~1 in the Main Text, in a Dynamical Casimir configuration. The solid curve is a Gaussian fit of the form
\begin{equation}\label{eq:fit}
 f(\Delta x)=\alpha+\beta\,e^{-(\Delta x-\zeta)^2/2\delta^{2}},
\end{equation}
whose parameters $\alpha$, $\beta$, $\delta$, and $\zeta$ have been determined through a standard non-linear chi-squared fit procedure, yielding
\begin{equation}\label{eq:fitpar}
 \alpha\approx0.0196,\quad
 \beta\approx-0.146,\quad
 \delta\approx15.6\xi_{-},\quad
 \zeta\approx117\,\xi_{-}.
\end{equation}
The presence of entanglement is attested by the negative dip centered around $\Delta x=\zeta\approx117\xi_{-}$. The position of the dip is in good agreement with the separation $\delta x(t_{\rm obs})=2|v^{\rm gr}(\tilde k)|t_{\rm obs}\approx120\xi_{-}$ that is expected for the ballistic propagation of phonons at $\tilde k=0.25/\xi_{-}$. As described in the Main Text, the width $\Delta\approx15.6\,\xi_{-}$ of the negative dip is determined by the spatial width of the wavepackets, $\sigma=15\,\xi_{-}$.

To quantify the numerical precision with which the quadrature variances have to be measured to assess entanglement, we quote here the values of the full covariance matrix at the two points indicated in the figure as open and filled dot. The former corresponds to the distance $\Delta x=120\xi_{-}$ where entanglement is strongest and the covariance matrix is  
\begin{equation}
V_{120}= 
\begin{pmatrix}
 0.624 & 0 & 0.210 & -0.284 \\
 0 & 0.624 & -0.284 & -0.210 \\
 0.210 & -0.284 & 0.624 & 0 \\
 -0.284 & -0.210 & 0 & 0.624
\end{pmatrix}.
\end{equation}
The latter corresponds to a distance $\Delta x=240\xi_{-}$ at which the the phonon wavepacket operators are effectively uncorrelated. As a result, the submatrix $C$ of the covariance matrix defined in Eq.~(2) of the Main Text is zero and moreover the full covariance matrix is diagonal:
\begin{equation}
V_{240} = 0.624\;{\mathbf I}_{4\times4},
\end{equation}
where ${\mathbf I}_{4\times4}$ is the $4\times4$ identity matrix. All matrix elements smaller than $10^{-5}$ have been set to zero.

\section{A possible experimental implementation}

In this Section, we provide more details on strategies to experimentally assess the entanglement by evaluating the generalized Peres-Horodecki function ${\mathcal P}$ defined in Eq.~(1) of the Main Text. To this purpose, the crucial step is a measurement of the quadratures of the phonon wavepacket mode operators defined in Eq.~(4) of the Main Text and entering into the definition of ${\mathcal P}$. 

Here, we shall see how these quadratures can be extracted from non-destructive measurements of a suitable Fourier component of the atomic density. 
A promising way to perform this measurement is to put the condensate into an optical cavity and measure the frequency shift of the cavity mode due to the condensate: as it was demonstrated in~\cite{esslingerscience,BECincavity,cavity_review}, this frequency shift is in fact proportional to the amplitude of a given phonon mode. The selected phonon mode is fixed by the spatial profile of the cavity mode. 

We begin our discussion from the simplest case of a homogeneous and stationarily flowing one-dimensional Bose-condensed atomic gas. In a naive Bogoliubov picture~\cite{BECbook,bectextbook}, the atomic field can be decomposed as a classical coherent condensate and quantized phonon excitations on top of the condensate:
\begin{multline}
 \hat\Psi(t,x)=e^{-i(\Omega t-K x)}\left[\Phi_0+\right.\\ \left.
 +\int \frac{dk}{2\pi}\left(\hat a_k u_k e^{-i(\om_k t-kx)}+\hat a_k^\dagger v_k^* e^{i(\om_k t-kx)}\right) \right] ,
\end{multline}
where $K$ and $\Omega=\hbar K^2/2m + gn$ are the momentum and the frequency of the condensate, respectively, $\Psi_0$ is chosen as positive and $u_k$ and $v_k$ as real. Assuming that the occupation numbers of phonon modes are small enough for the non-condensed fraction to be much less than unity, the atomic density operator can be written in the Bogoliubov form as
\begin{multline}
 \hat n(t,x) =\hat \Psi^\dagger(t,x)\,\hat \Psi(t,x) \\
 \approx \Phi_0^2 + \Phi_0\int \frac{dk}{2\pi}(u_k+v_k)\left(\hat a_k\, e^{-i(\om_k t-kx)}+\mbox{h.c}
 \right),
\end{multline}
where h.c. stands for Hermitian conjugate.

As discussed in~\cite{esslingerscience,BECincavity}, the optomechanical coupling between the cavity mode and the atomic gas can be written in the form
\begin{equation}
\mathcal{\hat H}_{\rm int}= \frac{g^2}{\Delta}\,\int dx\,\mathcal{\hat{E}}^\dagger(x)\,\mathcal{\hat{E}}(x)\,\hat{n}(x).
\label{eq:Hint}
\end{equation}
where $\hat{\mathcal E}(x)=\mathcal{E}_c(x)\,(\hat{c}+\hat{c}^\dagger)$ is the electic field operator in the (single-mode) cavity mode at position $x$ , $g$ is the atom-light coupling and $\Delta$ is the detuning between the atomic transition and the cavity mode frequency.
Assuming that the light field in the cavity forms a standing wave of amplitude $\mathcal{E}_c(x)=A \cos(k_0 x/2)$, the interaction Hamiltonian involves the overlap 
\begin{multline}
 {\mathcal{\hat O}}_{k_0}=
 2\int dx\, \hat n(x) \cos^2\left(\frac{k_0 x}{2}\right) \\	
 = \int dx\, \hat n(x) \left[1+\cos(k_0 x)\right],
\end{multline}
where the first constant term in square brackets is proportional to the number $\Phi_0^2 V$ of condensate atoms in the interaction region and the second term is associated to a combination $\hat{n}_{k_0}$ of phonon operators with wavevector $\pm k_0$ of the form:
\begin{multline}\label{eq:X+Xold}
\hat n_{k_0}=\frac{(u_{k_0}+v_{k_0})\Phi_0}{2}\left(\hat{a}_{k_0}+\hat{a}^\dagger_{k_0}+\hat{a}_{-k_0}+\hat{a}^\dagger_{-k_0}\right) = \\
=\frac{(u_{k_0}+v_{k_0})\Phi_0}{\sqrt{2}}(\Xq + \Xmq).
\end{multline}
Here, we have used $u_{k_0}=u_{-k_0}$ and $v_{k_0}=v_{-k_0}$ and we have defined the phonon quadrature operators as $\Xq=(\hat{a}_{k_0}+\hat{a}^\dagger_{k_0})/\sqrt{2}$ and $\Xq=(\hat{a}_{k_0}-\hat{a}^\dagger_{k_0})/\sqrt{2}i$.

Noting that the interaction Hamiltonian \eqref{eq:Hint} is of the generic form
\begin{equation}
 \mathcal{\hat H}_{\rm int}=G\,\hat{c}^\dagger\,\hat{c}\,\hat{\mathcal{O}}_{k_0}
\end{equation}
in terms of the cavity mode operator $\hat{c}$, a non-destructive measurement of the overlap operator $\hat{\mathcal{O}}_{k_0}$ can be obtained by an estimate of the cavity frequency shift, which in turn can be extracted from a measurement of the transmission spectrum through the cavity.

In the states of the fluid that we are considering in the Main Text, the expectation value of $\hat{n}_{k_0}$ vanishes, so the expectation value of the overlap operator $\mathcal{\hat O}_{k_0}$ reduces to the condensate contribution
\begin{equation}
 \ev{\mathcal{\hat O}_{k_0}}=\Phi_0^2 V.
\end{equation}
As a result, $\hat n_{k_0}$ can be measured at any realization as
\begin{equation}
 \hat n_{k_0}= \mathcal{\hat O}_{k_0}-\ev{\mathcal{\hat O}_{k_0}}.
\end{equation}

By performing a similar measurement at a later time by $\Delta t$, using a detector spatially displaced by $\Delta x$, different combinations of the operators $\hat a_{k_0}$, $\hat a_{k_0}^\dagger$, $\hat a_{-k_0}$, and $\hat a_{-k_0}^\dagger$ can be measured:
\begin{multline}\label{eq:nq}
 \hat n_{k_0}(\Delta t,\Delta x) \equiv \int dx\, n(\Delta t,x) \cos[k_0 (x-\Delta x)]\\
 =\frac{(u_{k_0}+v_{k_0})\Phi_0}{2}\left[
  \hat a_{k_0} e^{-i(\om_{k_0}\Delta t-{k_0}\Delta x)}
 \right.\\ \left.
 +\hat a_{-k_0} e^{-i(\om_{-{k_0}}\Delta t+{k_0}\Delta x)}
+\mbox{h.c.}
 \right],
\end{multline}
where
\begin{equation}
 \om_{k_0} = \tilde\om_{{k_0}}+v{k_0},\qquad\om_{-{k_0}} = \tilde\om_{{k_0}}-v{k_0},
\end{equation}
$\tilde\om_{k_0}$ is the phonon frequency in the reference frame where the condensate is at rest and $v={\hbar K}/{m}$
is the velocity of the condensate. By defining the spatial shift of the interference pattern in the comoving reference frame,
\begin{equation}
 \Delta \tilde x = \Delta x -v\Delta t,
\end{equation}
Eq.~\eqref{eq:nq} becomes
\begin{multline}
\hat n_{k_0}(\Delta t,\Delta \tilde x)
 =\frac{(u_{k_0}+v_{k_0})\Phi_0}{2}\left[
  \hat a_{k_0} e^{-i(\tilde\om_{k_0}\Delta t-{k_0}\Delta \tilde x)}
 \right.\\ \left.
 +\hat a_{-{k_0}} e^{-i(\tilde\om_{{k_0}}\Delta t+{k_0}\Delta \tilde x)}
+\mbox{h.c.}
 \right].
\end{multline}
We also define the temporal $\Delta\tau$ and spatial $\Delta\chi$ phase shift as
\begin{equation}
\Delta\tau\equiv \tilde\om_{k_0}\, {\Delta t} ,\qquad\Delta\chi\equiv {k_0}\,\Delta\tilde x.
\end{equation}
In terms of the normalized operator
\begin{equation}
 \Nq{\Delta\tau}{\Delta\chi}\equiv\frac{\sqrt{2}}{\Phi_0{(u_{k_0}+v_{k_0})}}\hat n_{k_0}(\Delta t,\Delta\tilde x).
\end{equation}
$\Xq+\Xmq$ of Eq.~\eqref{eq:X+Xold} reads
\begin{equation}\label{eq:X+X}
 \Xq+\Xmq = \Nq{0}{0}.
\end{equation}
and similarly
\begin{align}
 &\Xq-\Xmq = \Nq{\pit}{\pit},\label{eq:X-X}\\
 &\Pq+\Pmq = \Nq{\pit}{0},\label{eq:P+P}\\
 &\Pq-\Pmq = \Nq{0}{-\pit}.\label{eq:P-P}
\end{align}
Since $\Xq+\Xmq$ and $\Xq-\Xmq$ commute, the corresponding normalized phonon operators at different times $\Nq{0}{0}$ and $\Nq{\pit}{\pit}$ can be measured together on the same realization of the experiment. This allow to compute
\begin{align}
 &\Xq = \frac{1}{2}\left[\Nq{0}{0}+\Nq{\pit}{\pit}\right],\\
 &\Xmq = \frac{1}{2}\left[\Nq{0}{0}-\Nq{\pit}{\pit}\right]
\end{align}
for each realization of the experiment. After averaging over many realizations, one can estimate the expectation values $\ev{\Xq}$ and $\ev{\Xmq}$ as well as the variances $\ev{\Xq^2}$, $\ev{\Xmq^2}$, and the cross product
\begin{equation}\label{eq:XXm}
 \ev{\Xq\Xmq}=\frac{1}{4}\left[\ev{\Nq{0}{0}^2}-\ev{\Nq{\pit}{\pit}^2}\right].
\end{equation}
Note that, differently from a standard {\em in situ} measurement of the density via, e.g., phase contrast imaging~\cite{ketterle_sound} which simultaneously projects the many-body wavefunction on eigenstates of the density operators  $\hat{n}(x)$ at all positions $x$, our proposed cavity-assisted measurement is only sensitive to desired quadratures of the $\pm k_0$ momentum phonons and leaves all other quadratures and all other phonon modes unaffected. This is crucial to be able to perform a second measurement after a certain time $\Delta t$ without being disturbed by the result of the first measurement.

Analogously the operators $\Pq+\Pmq$ and $\Pq-\Pmq$ commute, so one can measure $\Nq{\pi/2}{0}$ and $\Nq{0}{-\pi/2}$ on the same realization. This allows to construct the expectation values $\ev{\Pq}$ and $\ev{\Pmq}$ as well as all required variances $\ev{\Pq^2}$, $\ev{\Pmq^2}$, and cross products
\begin{equation}\label{eq:PPm}
 \ev{\Pq\Pmq}=\frac{1}{4}\left[\ev{\Nq{\pit}{0}^2}-\ev{\Nq{0}{-\pit}^2}\right].
\end{equation}

The measurement of the remaining covariance $\ev{\Xq\Pq+\Pq\Xq}$ requires a bit longer procedure. As the commutators
\begin{multline}
 \comm{\Xq+\Xmq}{\Pq-\Pmq}=\comm{\Xq-\Xmq}{\Pq+\Pmq}\\
 =\comm{\Pq+\Xmq}{\Xq+\Pmq}=0
\end{multline}
vanish, all following quantities can be experimentally measured
\begin{align}
  &\ev{{\Nq{0}{0}}{\Nq{0}{-\pit}}}
  =\ev{\left(\Xq+\Xmq\right)\left(\Pq-\Pmq\right)}\nonumber\\
 &\qquad=\ev{\Xq\Pq}-\ev{\Xmq\Pmq}-\ev{\Xq\Pmq}+\ev{\Xmq\Pq},\\
 &\ev{{\Nq{\pit}{\pit}}{\Nq{\pit}{0}}}
  =\ev{\left(\Xq-\Xmq\right)\left(\Pq+\Pmq\right)}\nonumber\\
 &\qquad=\ev{\Xq\Pq}-\ev{\Xmq\Pmq}+\ev{\Xq\Pmq}-\ev{\Xmq\Pq},\\
 &\ev{\Nq{\pif}{-\pif}\Nq{\pif}{\pif}}
 =\ev{\left(\Pq+\Xmq\right)\left(\Xq+\Pmq\right)}\nonumber\\
 &\qquad=\ev{\Pq\Xq}+\ev{\Xmq\Pmq}+\ev{\Xmq\Xq}+\ev{\Pq\Pmq}.\label{eq:third}
\end{align}
Summing the first two equations, one obtains:
\begin{multline}\label{eq:sum}
 \ev{\Xq\Pq}-\ev{\Xmq\Pmq}
 =\frac{1}{2}\left[\ev{\Nq{0}{0}\Nq{0}{-\pit}}+\right. \\
 +\left.\ev{\Nq{\pit}{\pit}\Nq{\pit}{0}}\right].
\end{multline}
The remaining cross-correlation $\ev{\Xq\Pq+\Pq\Xq}$ can be then obtained by summing Eq.~\eqref{eq:sum} with Eq.~\eqref{eq:third}. In doing this, one has to note that $\comm{\Xq}{\Xmq}=0$ and use Eqs.~\eqref{eq:XXm} and~\eqref{eq:PPm}. The resulting explicit expression for the remaining cross-correlation finally reads:
\begin{widetext}
\begin{multline}
\ev{\frac{\Xq\Pq+\Pq\Xq}{2}}=
 \frac{1}{2}\ev{\Nq{\pif}{-\pif}\Nq{\pif}{\pif}}
 +\frac{1}{4}\left[\ev{\Nq{0}{0}\Nq{0}{-\pit}}+\ev{\Nq{\pit}{\pit}\Nq{\pit}{0}}\right]\\
 +\frac{1}{8}\left[\ev{\Nq{\pit}{\pit}^2}+\ev{\Nq{0}{-\pit}^2}-\ev{\Nq{0}{0}^2}-\ev{\Nq{\pit}{0}^2} \right].
\end{multline}
\end{widetext}
In conclusion, with this lengthy discussion we have shown how a suitable series of measurements of the atomic density provides full information on all the variances of the quadratures of a given phonon mode in a homogeneous atomic gas.

\begin{figure}
 \includegraphics[width=8cm]{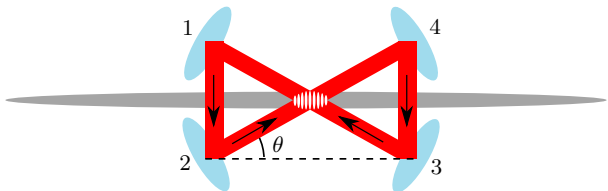}
 \caption{\label{fig:exp}Sketch of the proposed experimental cavity configuration. The cavity mode has a ring structure with two intersecting running waves generating a longitudinal standing wave pattern in the region of interest.}
\end{figure}

This protocol is straightforwardly generalized to a spatially-selective measurement of the quadratures of phonon wavepacket operators by taking into account the finite waist of the cavity mode $\sigma_c$. The phonon quadratures at different spatial positions $x_{1,2}$ that enter the gPH function $\mathcal{P}$ can be measured using two independent cavity modes localized at different positions: commutation of observables at different positions is guaranteed provided $|x_1-x_2|\gg\sigma_c$, as shown in the Main Text.
Remarkably, this procedure provides exactly the same information obtained from the wavepacket operators Eq.~(4) extracted in the numerical simulations from the convolution of the atomic field $\hat\Psi$ with Gaussian wavepackets. 

A possible experimental implementation of this procedure is depicted in Fig.~\ref{fig:exp}, where the cavity mode is enclosed by four mirrors. At the condensate position, the cavity mode consists of a standing wave created by the interference of two intersection fields forming an angle $\theta$ with the condensate velocity. Taking into account the waist $\sigma_c$ of the cavity mode, the longitudinal profile of the cavity mode in this region has the form 
\begin{equation}
 \mathcal{E}_c(x)=A \cos\left(\frac{k_0 x}{2} \right) e^{-(x-x_0)^2/2\sigma},
\end{equation}
where $k_0=2 k_c \cos\theta$, $k_c$ is the wavevector of the field forming the standing wave in the cavity, and $\sigma=\sigma_c/\sin\theta$.

Note that in this apparatus, the spatial shift $\Delta x$ is easily controlled by changing the relative phase between the two beams by, for instance, varying the optical paths from mirror 1 to mirror 2 and from mirror 3 to mirror 4, while keeping the total path constant.

\end{document}